\documentclass[aps,prb,preprint,groupedaddress]{revtex4}

\usepackage{amsmath}
\usepackage{graphicx}
\usepackage{dcolumn}
\usepackage{bm}
\usepackage{subfigure}

\begin{document}

\title{Zero-temperature generalized phase diagram of the \\ 
4$d$ transition metals under pressure}

\author{C. Cazorla$^{1,2,3}$}
\author{D. Alf\`e$^{1,2,3,4}$}
\author{M. J. Gillan$^{1,2,3}$}
\affiliation{$^{1}$London Centre for Nanotechnology, UCL, 
London WC1H OAH, U.K. \\ 
$^{2}$Department of Physics and Astronomy, UCL, London WC1E 6BT, U.K.\\
$^{3}$Materials Simulation Laboratory, London WC1E 6BT, U.K. \\
$^{4}$Department of Earth Sciences, UCL, London, WC1E 6BT, U.K.}

\begin{abstract}
We use an accurate implementation of density functional
theory (DFT) to calculate the zero-temperature 
generalized phase diagram
of the 4$d$ series of transition metals from Y to Pd as a function
of pressure $P$ and atomic number $Z$. The implementation used
is full-potential linearized augmented plane waves (FP-LAPW), and
we employ the exchange-correlation functional recently developed
by Wu and Cohen. For each element, we obtain the ground-state
energy for several crystal structures over a range of volumes,
the energy being converged with respect to all technical parameters
to within $\sim 1$~meV/atom.
The calculated transition pressures for all the elements and all
transitions we have found are compared with experiment wherever
possible, and we discuss the origin of the significant
discrepancies. Agreement with experiment for the zero-temperature
equation of state is generally excellent. The generalized phase diagram
of the 4$d$ series shows that the major boundaries slope
towards lower $Z$ with increasing $P$ for the early elements, as expected from the
pressure induced transfer of electrons from $sp$ states to
$d$ states, but are almost independent of $P$ for the later elements.
Our results for Mo indicate a transition
from bcc to fcc, rather than the bcc-hcp transition expected
from $sp$-$d$ transfer.
\end{abstract}

\maketitle

\section{Introduction}
\label{sec:introduction}

The transition metals are among the most intensively studied
families of elements, and there has been an effort going back
several decades to map and interpret systematic trends in their
properties. Some of these trends are of fundamental importance
to our understanding of the energetics and electronic structure
of transition metals. Examples include: the parabolic dependence of
cohesive energy and bulk modulus on $d$-band 
filling;~\cite{friedel69} the sequence of
most stable crystal structures associated with increasing band filling
at ambient pressure;~\cite{duthie77} and the pressure induced increase of
$d$-band width going roughly as the fifth inverse power of the
atomic volume.~\cite{heine67} However, our knowledge of transition-metal
systematics is still very far from complete, as is evident from the
current major controversies over high-pressure 
melting curves.~\cite{cazorla07,taioli07,errandonea05,ross07}
Even at low temperatures, there are sizable gaps in the map of transition-metal
phase diagrams, and new crystal structures continue to be
discovered.~\cite{ding07} We report here a systematic investigation of
the zero-temperature phase diagram of all the 4$d$ transition metals
over a wide range of pressures,
based on one of the most accurate implementations
of density functional theory (DFT) currently available.

There have, of course, been very many previous detailed studies
of transition metals based on DFT, some of which investigated
the relative stability of different crystal structures at high
pressures. However, most previous work has been designed to
address particular questions relating to particular metals. Here,
by constrast, the aim is to obtain a coherent overall view of
an entire transition-metals series. Specifically, using $Z$ to denote
atomic number, we wish to use DFT to map out the generalized $( P, Z )$
phase diagram of the 4$d$ series at $T = 0$~K at pressures $P$ up
to $\sim 500$~GPa~\cite{generalized}. In order 
to substantiate the accuracy of the
calculations, we shall compare our calculated results for $P$ as
a function of volume $V$ with all available experimental data.
The calculations are all performed using the
FP-LAPW implementation of DFT (full-potential linearized augmented
plane waves),~\cite{andersen75,koeling75,koeling77,singh94} which for a given exchange-correlation functional
is among the most accurate ways of calculating the total energy.
We use the exchange-correlation functional recently developed
by Wu and Cohen,~\cite{wu06} which appears to reproduce the experimental
energetics of 4$d$ transition metals more accurately than other
functionals.~\cite{tran07}

There are several motivations for wishing to obtain the $( P, Z )$
phase diagram of a transition-metal series at $T = 0$~K.
One motivation is the possibility of discovering hitherto
unknown crystallographic phase boundaries for particular metals.
Another motivation is to probe our basic understanding of transition-metal
energetics. At ambient pressure, the most stable crystal structures
of the 4$d$ and 5$d$ series follow the sequence
hexagonal close packed (hcp) to body-centred cubic (bcc) to hcp to
face-centred cubic (fcc) with increasing $Z$, and the 3$d$ series
nearly follows this sequence, the deviations being caused by
magnetic effects.~\cite{brewer67} This sequence is 
entirely explained by band energies,
and can be reasonably well reproduced using a canonical $d$-band
tight-binding model with some modifications due to hybridization with
sp bands.~\cite{pettifor77,dalton69,ducastelle71} The main effect of pressure is 
to shift the relative energies
of the $d$-band centroid and the bottom of 
the $sp$ band, and hence increase the
$d$-band filling.~\cite{pettifor77} This leads one to expect that the same sequence
of structures will be found at high $P$, but with the boundaries
shifted to the left (towards lower $Z$)~.~\cite{moriarty92,mcmahan86} We 
shall confirm that this is the case, though the reality is slightly
more complicated.

In addition to the motivations we have mentioned, there is another
that is important to us. Current controversies over the high-pressure
melting curves of transition metals stem from apparent disagreements
between melting temperatures deduced from static compression and shock
experiments, these disagreements amounting to several thousand K
at megabar pressures.~\cite{errandonea05,ross07} DFT calculations of transition-metal
melting curves~\cite{cazorla07,taioli07,belonoshko04} support 
the correctness of the shock results, and one
of the proposed resolutions of the conflicts is that the
transitions identified as melting in static compression are in fact
solid-solid transitions. It would clearly help to reduce the confusion
if we had a better understanding of the generalized $( P, T, Z )$ phase diagram
of transition-metal series~\cite{generalized}. At 
present, we have a fairly complete
knowledge of this only at low $P$, and, to a lesser extent, at 
low $T$. We see the present attempt to complete the low-$T$
$( P, Z )$ diagram as an essential step towards mapping the
full $( P, T, Z )$ diagram.

Earlier DFT work has already given much information about the
energetics of transition metals at $T = 0$~K.
Particularly important here is the work of 
Pettifor~\cite{pettifor77,pettifor70,pettifor72} and others,
which provided a systematic understanding of the structural
trends across all the transition-metal series in terms
of the electronic densities of states of different
crystal structures. Their work also gave important insights
into the pressure induced transfer of electrons from sp-states
to d-states. Also important was the first-principles
work of Skriver~\cite{skriver85} on structural trends across transition
metal series at zero pressure. The pioneering work of Moriarty, Johansson
and others on the high-pressure energetics of transition
metals will be cited below.
The information from all this previous
work could be assembled to produce a substantial part of the 
$( P, Z )$ diagram, but there would be inconsistencies from the
use of different DFT implementations and exchange-correlation
functionals. Here, we avoid such inconsistencies by using a 
single high-accuracy method for all the calculations.

The remainder of the paper is organized as follows. 
In Sec.~\ref{sec:techniques},
we summarize the essential ideas underlying the methods 
used in this work and the results
of our convergence tests. Then (Sec.~\ref{sec:results}), we present for each 
of the transition metals from Y to Pd our
calculated energy differences between different crystal
structures and the equation of state (EOS) $P ( V )$ 
for pressures from ambient to $\sim 500$~GPa, 
comparing our results with experiments and previous calculations
where possible. At the end of Sec.~\ref{sec:results}, we 
present the zero-temperature $( P, Z )$ phase diagram of
the 4$d$ transition-metal series.  
Discussion and conclusions are in Sec.~\ref{sec:discussion}.

\section{Techniques and tests}
\label{sec:techniques}

There are several implementations of DFT that can be used to
calculate the total energy per atom of a crystal, including
pseudopotential techniques,~\cite{vanderbilt90} the projector augmented wave
technique,~\cite{blochl94,kresse99} the full-potential augmented plane-wave (FP-LAPW)
technique,~\cite{andersen75,koeling75,koeling77,singh94} etc. We have chosen to use FP-LAPW here, because
for a given exchange-correlation functional it can 
give values for the total energy that are closer to the
exact value than most other implementations. There are a number of
technical parameters in FP-LAPW that control convergence towards the
exact value. We recall in this Section the main ideas of
FP-LAPW, summarize the parameters that control convergence,
and report tests that guide our setting of these parameters.
The exchange-correlation functional used for all the present work
is that due to Wu and Cohen.~\cite{wu06} Later in this Section, we outline briefly
what this functional is, and we report tests indicating that it
should be particularly accurate for present purposes.

\subsection{Full-potential linearized augmented plane waves}
\label{sec:FP-LAPW}

In APW methods, space is divided into non-overlapping muffin-tin
spheres (radius $R_{\rm MT}$) centred on the atoms, and the interstitial
region between the spheres. Each Kohn-Sham orbital is
represented as a sum of radial functions multiplied
by spherical harmonics up to a maximum angular momentum
$l_{\rm orb}^{\rm max}$ in each sphere, and as a sum of plane waves
(maximum wavevector $K_{\rm max}$) in the interstitial region;
the coefficients are adjusted to achieve continuity at the
sphere boundaries. The Kohn-Sham eigenstates are divided
into core and valence states; core states are treated as being
non-zero only inside the muffin-tin spheres, while valence
states extend over all space. The radial functions used as
basis sets are solutions of the Schr\"{o}dinger equation
inside the atomic spheres, and are therefore energy dependent.
In linearized APW (LAPW), the energy dependence is treated in
a linear approximation. 
In full-potential implementations (FP-LAPW),
the Kohn-Sham potential is also represented as a sum over
angular momentum functions within the spheres (maximum angular
momentum $l_{\rm pot}^{\rm max}$) and a sum over plane waves
(maximum wavevector $G_{\rm max}$) in the interstitial region.
As it stands, this scheme is not accurate if there are semi-core
states, i.e. low-lying states that are not adequately treated
by linearization. One solution to this problem is to linearize the
basis functions using different reference energies for valence
and semi-core states. Alternatively one can augment
the basis set with ``local orbitals'', this procedure being
known as the ``APW+lo'' method. We use the full-potential
version of the latter method here 
in order to treat all the valence and semi-core states. 
A fully relativistic treatment is
used for core states, and a scalar relativistic treatment for
all other states. The wavevectors ${\bf k}$ for which the Kohn-Sham
equations are self-consistently solved must be sampled over
the Brillouin zone of the crystal, and the well-known
Monkhorst-Pack $k$-point sampling scheme is used for this purpose.~\cite{monkhorst76}
All the calculations were performed using the WIEN2k code.~\cite{blaha01}

Throughout the present calculations, we aim to achieve
convergence of the total energy to within 1~meV/atom with
respect to all the parameters we have just mentioned.
All states having principal quantum number $n \le 3$ are treated
as core states, and all higher states as valence states.   
This choice is based on the fact that the states with $n = 3$ 
lie at least 200~eV
below the Fermi level so they will not respond significantly 
to compression even at megabar pressures.
Provided the muffin-tin spheres do not overlap, and provided the
calculations are fully converged with respect to all parameters,
the choice of $R_{\rm MT}$ should not affect the results, but
it does affect the efficiency of the calculations. We used the
default setting of $R_{\rm MT}$ provided by the WIEN2k algorithm,
which ensures that $R_{\rm MT}$ varies appropriately
as the volume per atom is varied. To determine
the settings required for the technical parameters
$K_{\rm max}$, $G_{\rm max}$, $l_{\rm orb}^{\rm max}$ and
$l_{\rm orb}^{\rm pot}$, and for $k$-point sampling, we have
conducted systematic tests on bcc, fcc and hcp Mo; we assume that the settings
that give the required accuracy for Mo will also serve for
all the 4$d$ transition metals in all the structures of interest. 

Our tests on Mo indicate that the choices $R_{\rm MT} K_{\rm max} = 9.5$,
$G_{\rm max} = 18$~bohr$^{-1}$, $l_{\rm orb}^{\rm max} = 10$ and
$l_{\rm pot}^{\rm max} = 4$, together with 
$14 \times 14 \times 14$ $k$-point sampling 
($14 \times 14 \times 7$ for hcp), give convergence
to within $\sim 1$~meV/atom. To demonstrate that this is the
case, we have set all of these parameters except one to the values
just quoted, and studied the effect of varying the free parameter.
The values of total energy per atom $E_{\rm tot}$ obtained
when we vary in turn $R_{\rm MT} K_{\rm max}$, $G_{\rm max}$,
$l_{\rm orb}^{\rm max}$, $l_{\rm pot}^{\rm max}$ and
$k$-point sampling are reported in Table~I~,
from which we see that the required convergence is indeed
achieved with the quoted values. These are the values used
for all the calculations reported later. We note that,
although we have paid careful attention to convergence,
there remain small errors due to the linearization inherent
in the FP-LAPW scheme. 

\begin{table}
\begin{center}
\label{tab:convergence}
\begin{tabular}{c c}
\hline
\hline
$k$-points      &         $ E_{\rm tot}$  \\
\hline
$10 \times 10 \times 10  $      &         $-110132.5640   $       \\
$14 \times 14 \times 14  $      &         $-110132.5669   $       \\
$16 \times 16 \times 16  $      &         $-110132.5658   $       \\
$18 \times 18 \times 18  $      &         $-110132.5664   $       \\
\hline
$G_{\rm max}$ (bohr$^{-1}$)     &         $E_{\rm tot}$   \\
\hline
$14  $          &         $-110132.5668   $       \\
$18  $          &         $-110132.5669   $       \\
$20  $          &         $-110132.5669   $       \\
\hline
$l_{\rm pot}^{\rm max}$         &         $E_{\rm tot}$  \\
\hline
$ 4 $   &         $-110132.5669  $        \\
$ 6 $   &         $-110132.5669  $        \\
\hline
$R_{\rm MT} K_{\rm max}$        &         $E_{\rm tot}$   \\
\hline
$ 8.0 $         &         $-110132.5250   $       \\
$ 9.0 $         &         $-110132.5613   $       \\
$ 9.5 $         &         $-110132.5669   $       \\
$ 10.5 $        &         $-110132.5671   $       \\
\hline
$l_{\rm orb}^{\rm max}$         &         $ E_{\rm tot} $  \\
\hline
$ 10 $          &         $-110132.5669  $        \\
$ 12 $          &         $-110132.5662  $        \\
\hline
\hline
\end{tabular}
\end{center}
\caption{Tests on convergence of FP-LAPW total energy per atom
$E_{\rm tot}$ (eV units) for bcc Mo. Variation of
$E_{\rm tot}$ is reported with respected to
$k$-point sampling, plane-wave cut-off $G_{\rm max}$
(bohr$^{-1}$ units) and angular momentum cut-off
$l_{\rm pot}^{\rm max}$ in representation of Kohn-Sham potential,
and plane-wave cut-off $K_{\rm max}$ and angular momentum
cut-off $l_{\rm orb}^{\rm max}$ in representation of orbitals
($R_{\rm MT}$ is muffin-tin radius). In each section of the Table,
only a single parameter is varied, the other parameters being
fixed as $14 \times 14 \times 14$ ($k$-point sampling),
$G_{\rm max} = 18$~bohr$^{-1}$, $l_{\rm pot}^{\rm max} = 4$,
$R_{\rm MT} K_{\rm max} = 9.5$ and $l_{\rm orb}^{\rm max} = 10$.
}
\end{table}

\subsection{Wu-Cohen exchange-correlation functional}
\label{sec:WC} 

The Wu-Cohen exchange-correlation functional is a particular form
of generalized gradient approximation (GGA). In GGAs, the
total exchange-correlation energy is the integral over the
volume of the system of a density of exchange-correlation energy,
this density being expressed in terms of the number density
$n ( {\bf r} )$ of electrons as 
$n ( {\bf r} ) \epsilon_{\rm xc} ( n ( {\bf r} ) , s ( {\bf r} ) )$.
Here, $\epsilon_{\rm xc}$, the exchange-correlation energy per
electron at position ${\bf r}$, depends not only on
$n ( {\bf r} )$ itself but also on the magnitude
of its dimensionless gradient
$s ( {\bf r} ) = | \nabla n | / [ 2 ( 3 \pi^2 )^{1/3} n^{4/3} ]$.
The energy $\epsilon_{\rm xc}$ is expressed as
$\epsilon_{\rm xc} ( n ( {\bf r} ) , s ( {\bf r} ) ) =
\epsilon_{\rm xc}^{\rm unif} ( n ( {\bf r } ) ) F_{\rm xc} ( s ( {\bf r} ) )$,
where $\epsilon_{\rm xc}^{\rm unif} ( n )$ is the exchange-correlation
energy per electron in the uniform electron gas of density $n$, and
$F_{\rm xc}$ is the so-called enhancement factor.

In the widely used Perdew-Burke-Ernzerhof (PBE) form of GGA,~\cite{perdew96} 
a parameter-free formula expressing
the dependence on $s$ of $F_{\rm xc} ( s )$ was derived by requiring
that certain exact conditions be satisfied. The WC functional
takes the same GGA form as PBE for the correlation part of $F_{\rm xc}$,
but modifies the exchange part. The modification was done with the
intention of improving the functional for condensed matter, at the
expense of a somewhat worse description of atoms and molecules.
Specifically, the modification was based on the idea that the
exchange part of $F_{\rm xc}$ should be constructed so as to
reproduce the fourth-order gradient expansion of the exact
exchange-energy functional of the electron gas in the limit
of slowly varying density.~\cite{svendsen96} The detailed form of the
exchange enhancement factor in the WC functional can be found
in the original paper,~\cite{wu06} which also presents the results of
tests suggesting that the functional gives better results
than LDA and PBE for the equilibrium lattice parameter and
bulk modulus of a range of materials. Subsequently, tests of WC
were reported on a much more extensive set of materials
using the FP-LAPW implementation of DFT.~\cite{tran07} These tests supported
the claim that WC is on average more accurate than LDA or PBE,
but they showed that the improvement over these other functionals
is by no means uniform. However, for 4$d$ transition metals at 
ambient pressure, the WC predictions of equilibrium lattice parameter
and bulk modulus are, with a few exceptions, better than both
PBE and LDA.

\begin{figure}
\centerline{
\includegraphics[width=0.8\linewidth]{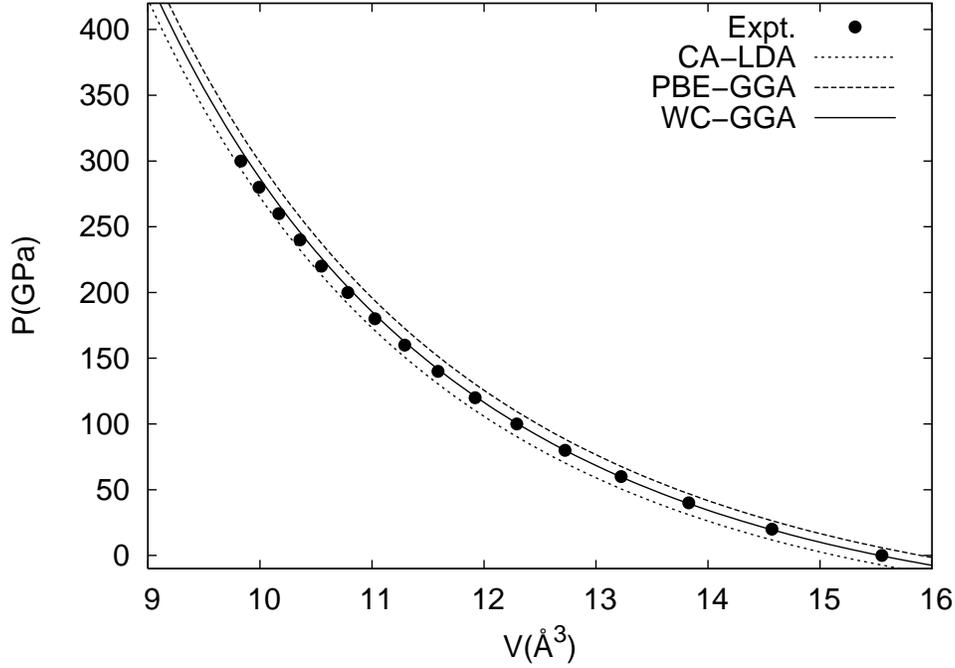}}%
\caption{Zero-temperature equation of state of Mo 
up to $P \sim 400$~GPa calculated with GGA(WC), GGA(PBE) and 
LDA(CA) exchange-correlation functionals. Dots 
show experimental data.~\cite{hixson92} }
\label{fig:wcpbeca}
\end{figure}

To confirm that the quality of WC is maintained at high pressures,
we compare in Fig.~\ref{fig:wcpbeca} our calculated results for pressure $P$ as
a function of volume $V$ at zero temperature for Mo obtained
with WC, LDA (Ceperley-Alder)~\cite{ceperley80} and PBE with experimental results
up to $P = 300$~GPa. The experimental measurements were performed
at room temperature but were corrected
for thermal effects so that they refer to $T = 0$~K.~\cite{hixson92} The
comparisons show that WC is in almost perfect agreement with
experiment over the entire pressure range, while both LDA and
PBE show significant deviations. Comparisons of calculated
$P ( V )$ curves with experimental data for other 4$d$ transition
metals presented in the next section further confirm the
quality of WC. 

\section{Results}
\label{sec:results}

For each of the 4$d$ transition metals from Y to Pd, we have
calculated the total energy at a closely spaced series of
atomic volumes for the body-centred cubic (bcc), face-centred cubic (fcc)
and hexagonal close-packed (hcp) structures. In some cases, we have
also studied other crystal structures. The pressure as a function
of volume is then obtained for each structure by fitting the
total energy results with a 3rd-order Birch-Murnaghan equation
of state.~\cite{birch47} From these results, we straightforwardly obtain the
most stable crystal structure at each pressure, as well as the pressures
of the structural transitions and the volumes of the coexisting phases
at these transitions. In the following, we present first for each
element the difference between the energy per atom in each
structure and the energy in bcc structure, as a function of $V$.
We then present $P ( V )$ of the stable structures over the
pressure range from ambient to typically 500~GPa. Where possible,
we compare $P ( V )$ and the transition pressures with experimental
measurements, and we indicate the relation with previous theoretical
results.

\subsection{Yttrium}
\label{sec:Y}

\begin{figure}
\centering
{ \includegraphics[width=0.65\linewidth]{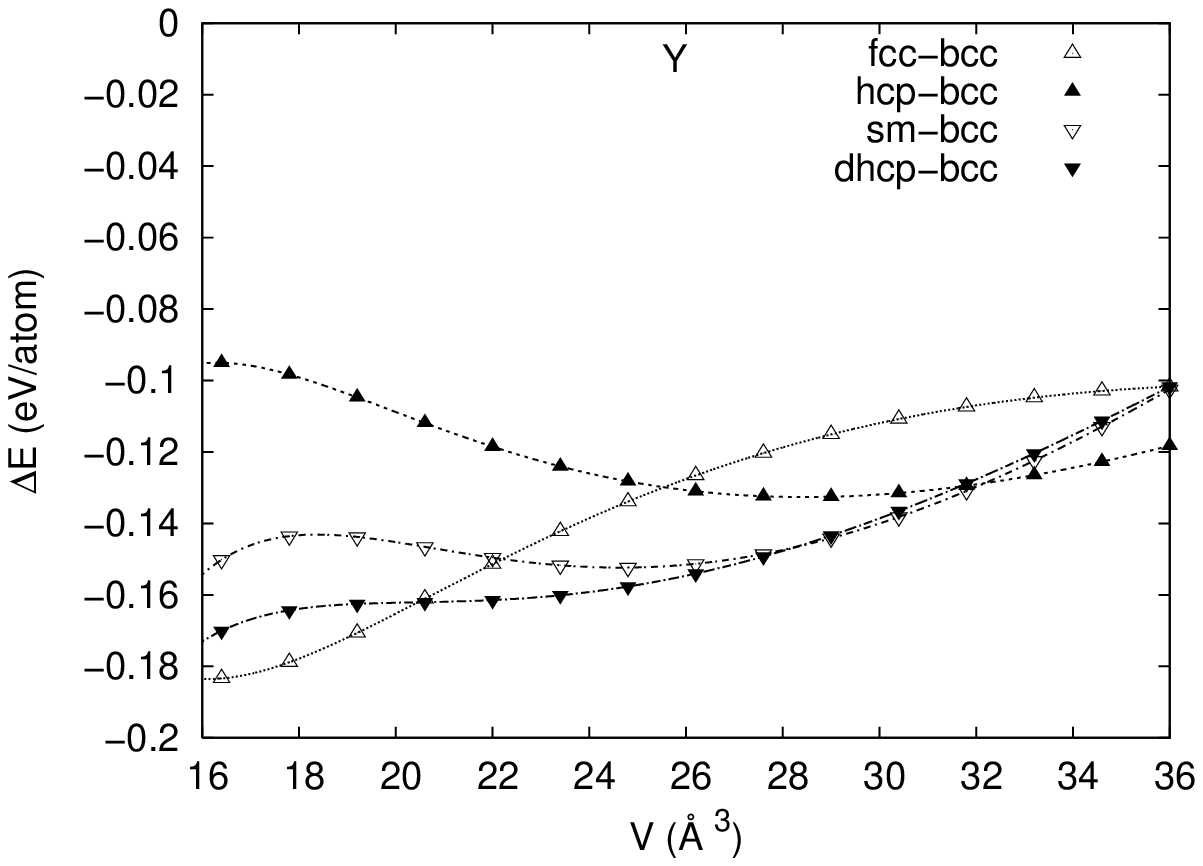} }%
{ \includegraphics[width=0.65\linewidth]{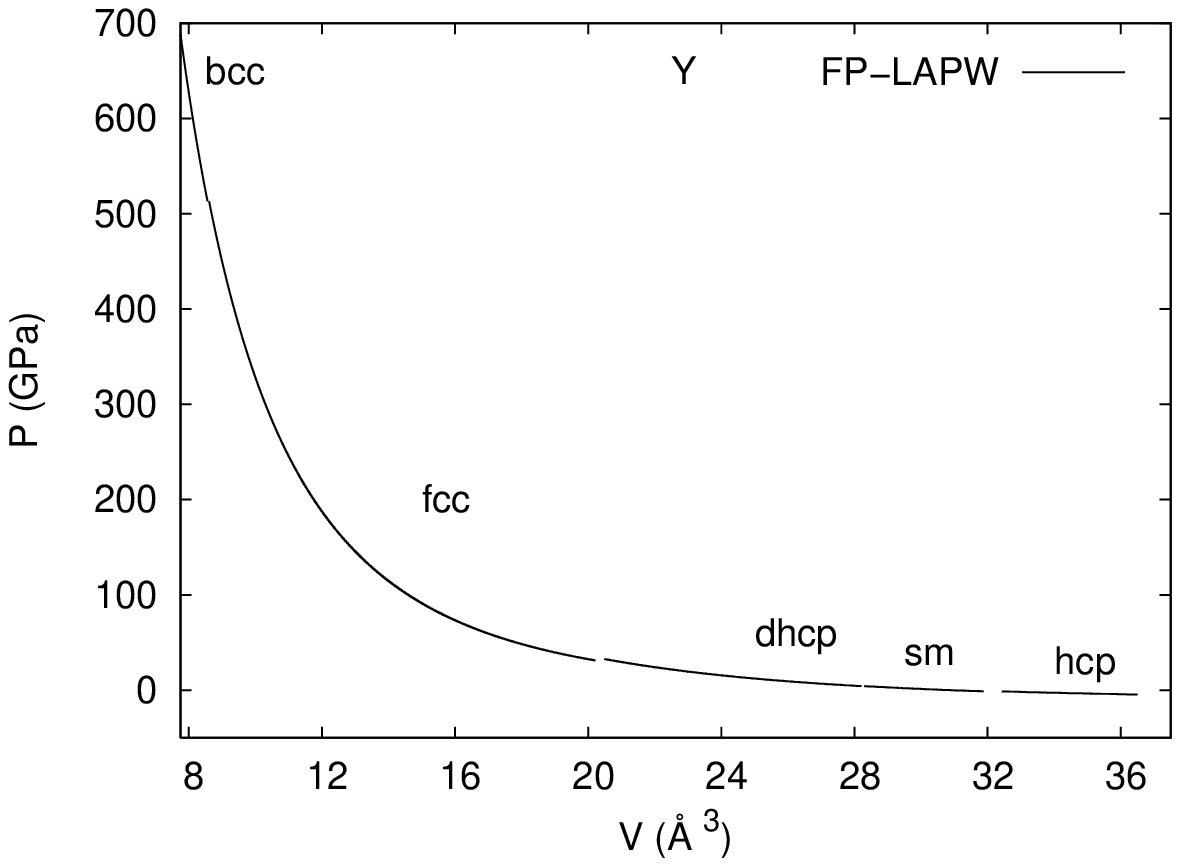} }%
\vspace{0.0cm}
\caption{\emph{Top}: Energy differences $\Delta E$ of the hcp, $\alpha$Sm (C19), 
dhcp (A3$^{\prime}$) and fcc structures with respect to bcc in Y as function of volume per atom.
\emph{Bottom}: Equation of state of Y up to $P \sim 700$~GPa. 
}
\label{fig:Y}
\end{figure}

\begin{table}
\begin{center}
\label{tab:Ypt}
\begin{tabular}{c | c c c c }
\hline
\hline
 transition \quad & $\qquad V_1 \qquad$ & $\qquad V_2 \qquad $ & \multicolumn{2}{c}{$\quad P_t$} \\
 & & & DFT & expt \\
\hline
 hcp - $\alpha$Sm & 32.44 & 31.88 & $-1.3$ ($-3^{a}$) & $\sim 15^{c} $ \\
 $\alpha$Sm - dhcp & 28.30  & 28.21 & 4.2 ($3^{a}$) & $\sim 30^{c}$ \\
 dhcp - fcc & 20.49 & 20.32 & 32.7 & $\sim 44^{d}$ \\
 fcc - bcc & 8.61 & 8.57 & 513 ($283^{b}$) & $-$ \\
\hline 
\hline
\end{tabular}
\end{center}
\caption{Calculated volumes $V_1$, $V_2$ (units of \AA$^3$/atom) and 
transition pressures $P_t$ (units of GPa) from present
work (earlier DFT results in parentheses), and experimental $P_t$
for structural phase transitions in Y.
Values from earlier work: a:~Lei {\em et al.}~\cite{lei07};
b:~Melsen {\em et al.}~\cite{melsen93}; c:~Vohra {\em et al.}~\cite{vohra81};
d:~Grosshans and Holzapfel~\cite{grosshans92}.
} 
\label{tab:Y}
\end{table}

Experimentally, Y has the hcp structure at ambient pressure, and
with increasing pressure passes successively to
the $\alpha$-samarium ($\alpha$Sm),
double hcp (dhcp) and fcc structures.~\cite{dhcp,vohra81,grosshans92}
This is the same sequence of structures observed in the lanthanides
under pressure, as has often been discussed.~\cite{johansson75}
Earlier DFT calculations on Y have correctly predicted these
structures,~\cite{lei07} and also indicated a
transition to bcc at very high pressure.~\cite{melsen93}
Fig.~\ref{fig:Y} reports our calculated energies of the hcp, $\alpha$Sm,
dhcp and fcc structures relative to bcc as a function of $V$.
This shows the sequence of stable structures in the experimentally
observed order. The resulting EOS is reported in Fig.~\ref{fig:Y}, and
the transition pressures and volumes are given in Table~II~.
Although the calculated sequence of crystal structures is the same as
that observed experimentally, we significantly underestimate the
transition pressures by at least 10~GPa. The difference is considerably
greater for the transition $\alpha$Sm~$\rightarrow$ dhcp, but it is
clear from Fig.~\ref{fig:Y} that the energy curves of these two structures
follow each other so closely that the prediction of this transition
pressure is likely to be very challenging. Possible reasons for
our underestimation of the transition pressures are discussed in 
Sec.~\ref{sec:discussion}~.

We also report in Table~\ref{tab:Y} the results of very recent
DFT work on Y up to $\sim 150$~GPa by Lei {\em et al.}~\cite{lei07},
as well as those of the older work of Melsen {\em et al.}~\cite{melsen93}.
The hcp-$\alpha$Sm and $\alpha$Sm-dhcp transition pressures
$P_t$ of Lei {\em et al.}, who used FP-LAPW based on LDA, are
within $2$~GPa of ours. It is known that some tranverse phonons of fcc Y are
unstable at pressures below $90$~GPa.~\cite{yin06} Based on this fact, 
Lei {\em et al.} analyze a distorted fcc structure which they find to be   
more stable than fcc. In order to save effort,  
however, we consider here the fcc structure directly. 
It is striking that the fcc-bcc $P_t$ of Melsen {\em et al.},
obtained from FP-LMTO calculations based on LDA is below ours
by over $200$~GPa. It appears to us that their treatment
of semi-core states was much more approximate than the one
used here, and this may account for the large difference.
The very small volume change of only $\sim 0.04$~\AA$^3$/atom
that we find for this transition will magnify the effect
of any technical errors in the prediction of $P_{t}$. We return to
this question in Sec.~\ref{sec:discussion}.

\subsection{Zirconium}
\label{sec:Zr}

\begin{table}
\begin{center}
\begin{tabular}{c | c c  c c c  c}
\hline
\hline
transition \quad & $\qquad V_{1} \qquad$ & $\qquad V_{2} \qquad $ & \multicolumn{3}{c}{$P_t$ (DFT)} & $\qquad P_t$ (expt) \\
  &  &  & WC & LDA & PBE & \\
\hline
hcp-$\omega$ & $25.47^a$ & $24.57^a$ & $-8^a$ & & & $3^e$ \\
$\omega$-bcc & $19.07^a$ & $18.43^a$ & $22^a$ & $24^b$ & $32^b$ & $30^f$ \\
             &           &           &        & $16^c$ & $28^c$ & \\
             &           &           &        & $5^d$  &        &  \\
\hline
\hline
\end{tabular}
\end{center}
\caption{Calculated volumes $V_1$, $V_2$ (this work, units
of \AA$^3$/atom) and calculated
and experimental transition pressures $P_t$ (units of GPa)
for structural transitions in Zr. Calculated $P_t$ values
are listed according to the exchange-correlation function
(WC, LDA or PBE) used. Sources: a:~present work; 
b:~Ostanin {\em et al.}~\cite{ostanin98}; c:~Schnell and Albers~\cite{schnell06};
d:~Jona and Marcus~\cite{jona03}; e:~Zhao {\em et al.}~\cite{zhao05};
f:~Hui {\em et al.}~\cite{hui90}. 
} 
\label{tab:Zr}
\end{table}

\begin{figure}
\centering
{\includegraphics[width=0.65\linewidth]{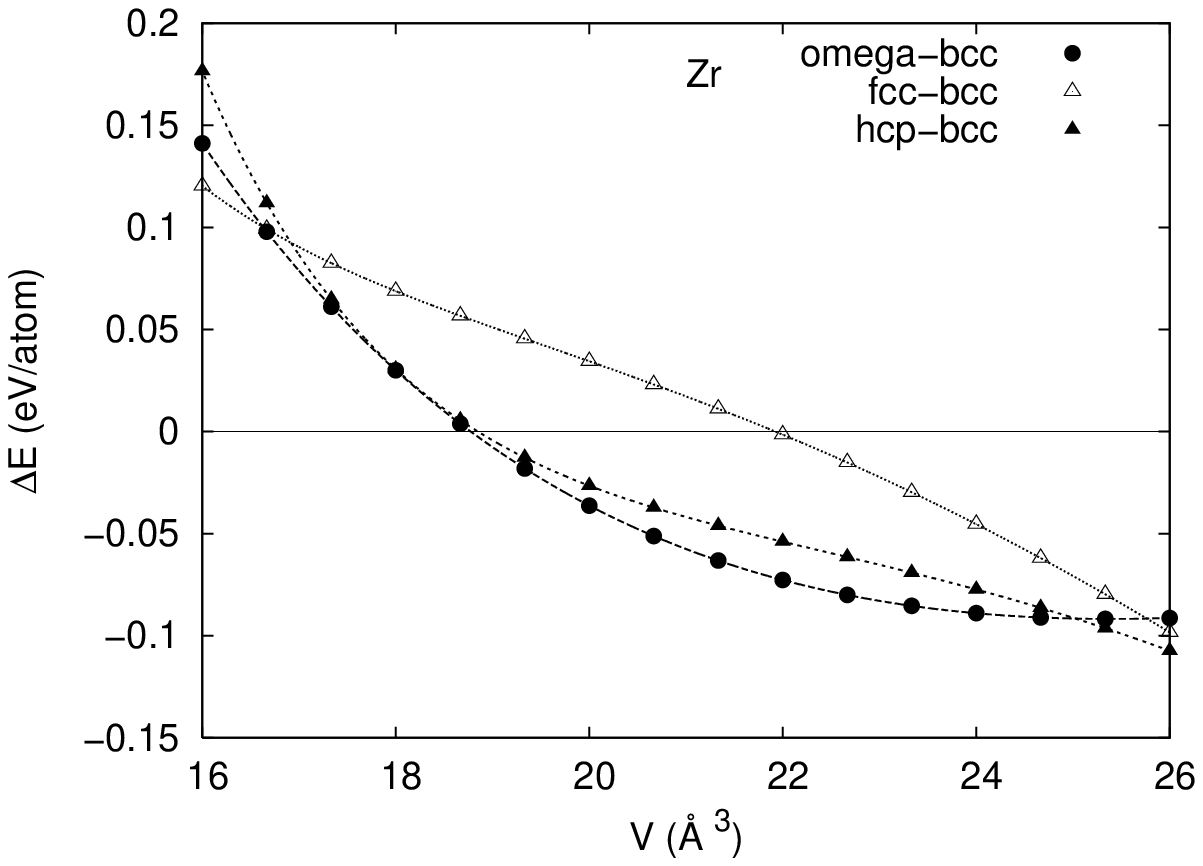} }%
{ \includegraphics[width=0.65\linewidth]{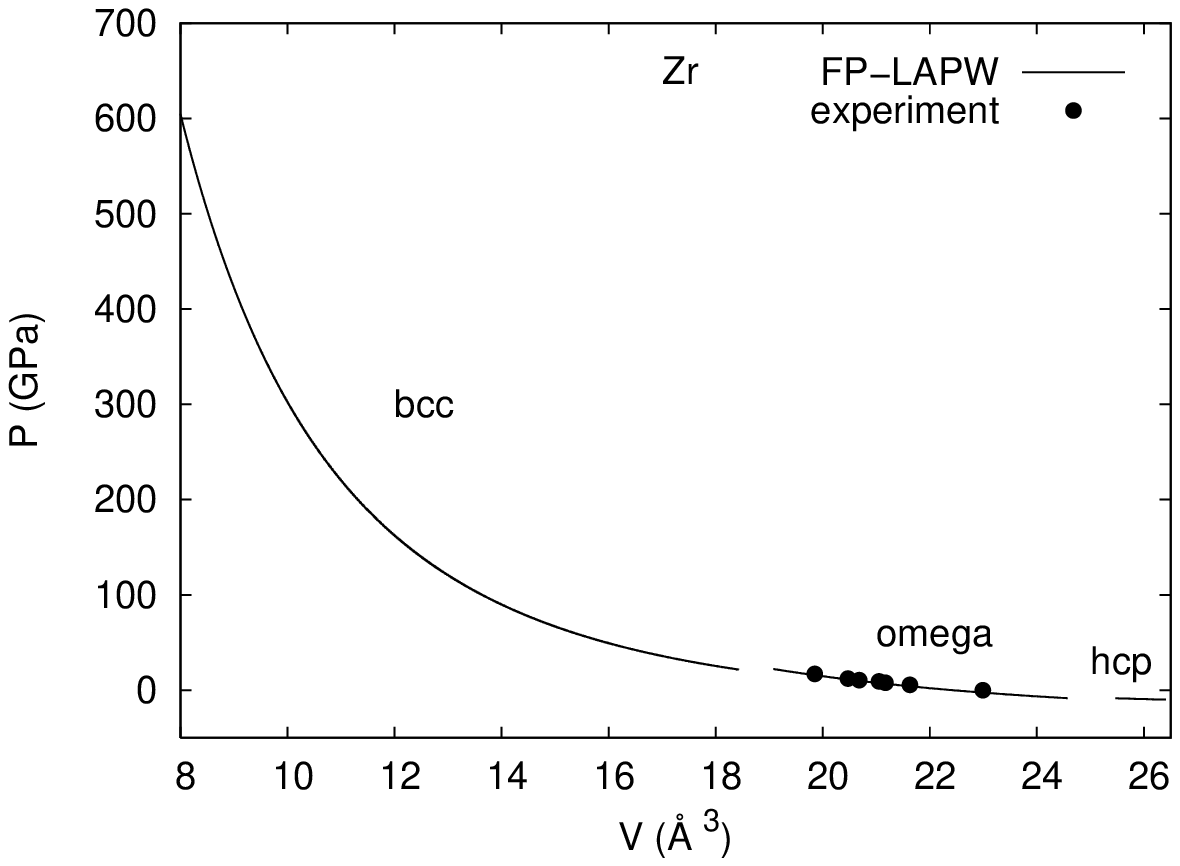} }%
\vspace{0.0cm}
\caption{\emph{Top}: Energy differences $\Delta E$ of the hcp, $\omega$ 
(C32) and fcc structures with respect to bcc in Zr as function of volume.
\emph{Bottom}: Equation of state of Zr up to $P \sim 600$~GPa; 
dots represent experimental data from Ref.~[\onlinecite{zhao05}]~. }
\label{fig:Zr}
\end{figure}

Experimentally, the stable structure of Zr at ambient
pressure is hcp, but measurements
show a transition to the $\omega$ phase at the rather low pressure
of $2.8$~GPa, followed by a transition to bcc at
$P \simeq 30$~GPa.~\cite{zhao05} Our calculated energies 
of the $\omega$, hcp and
fcc structures relative to bcc as a function of volume (Fig.~\ref{fig:Zr})
show the same sequence, and our calculated $P ( V )$ in
the $\omega$ structure is in close agreement with experimental data.
However, as in the case of Y, our calculations appear to underestimate
the transition pressures (see Table~III). We find that at
$P = 0$, $T = 0$, the $\omega$ structure is most stable, the transition
between hcp and $\omega$ occurring at the negative pressure
of $- 8.4$~GPa. Our calculated transition pressure between $\omega$
and bcc of $22.2$~GPa is also significantly below the experimental
value of $30(2)$~GPa.~\cite{hui90} The underestimation 
by $\sim 10$~GPa is similar to what we found for Y,
and will be discussed in Sec.~\ref{sec:discussion}.

Previous theoretical estimates of the $\omega$-bcc transition
pressure $P_{t}$ show considerable variation (Table~\ref{tab:Zr}).
It seems that $P_{t}$ from LDA is lower than that from PBE by
$\sim 10$~GPa, but there are significant differences between
results obtained with the same DFT functional. The results of
Jona and Marcus~\cite{jona03} seem to be seriously out of
line with both experiment and other calculations. We shall
comment further in Sec.~\ref{sec:discussion} 

\subsection{Niobium}
\label{sec:Nb}

\begin{figure}
\centering
{\includegraphics[width=0.65\linewidth]{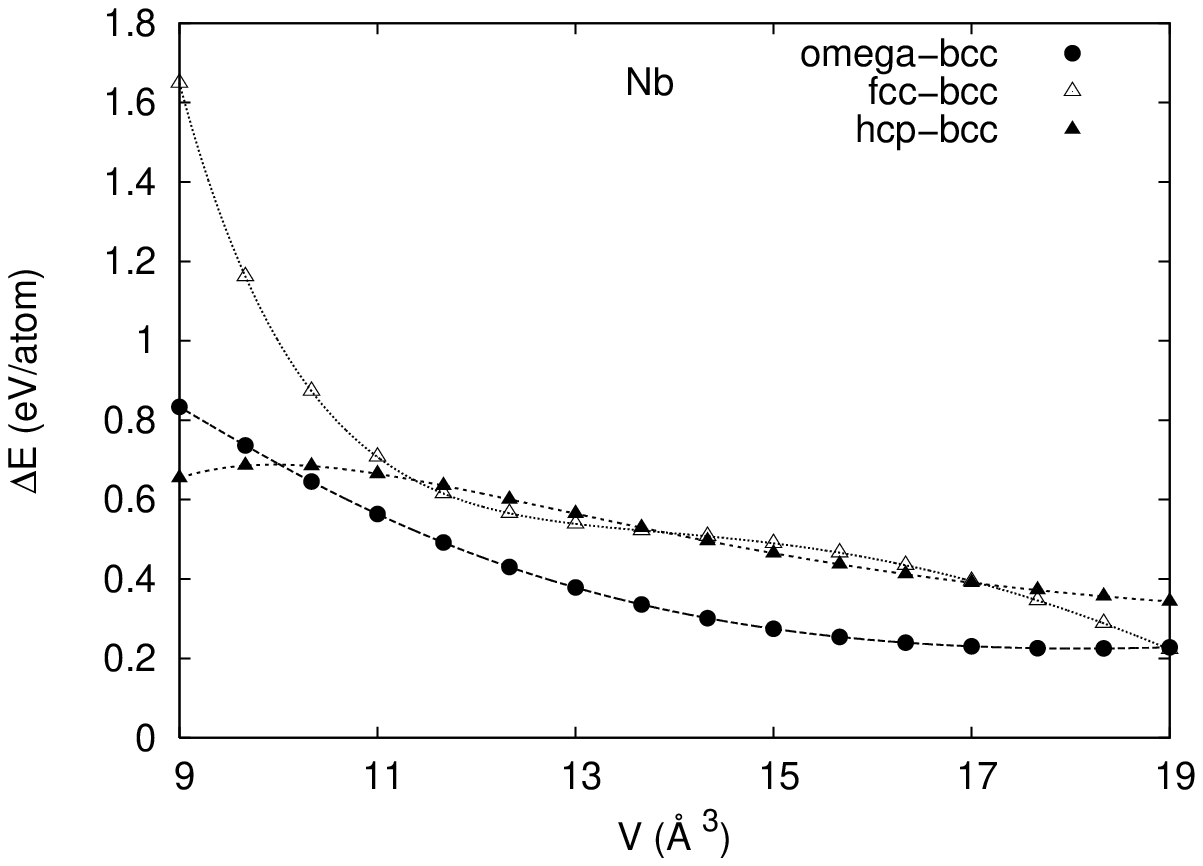} }%
{\includegraphics[width=0.65\linewidth]{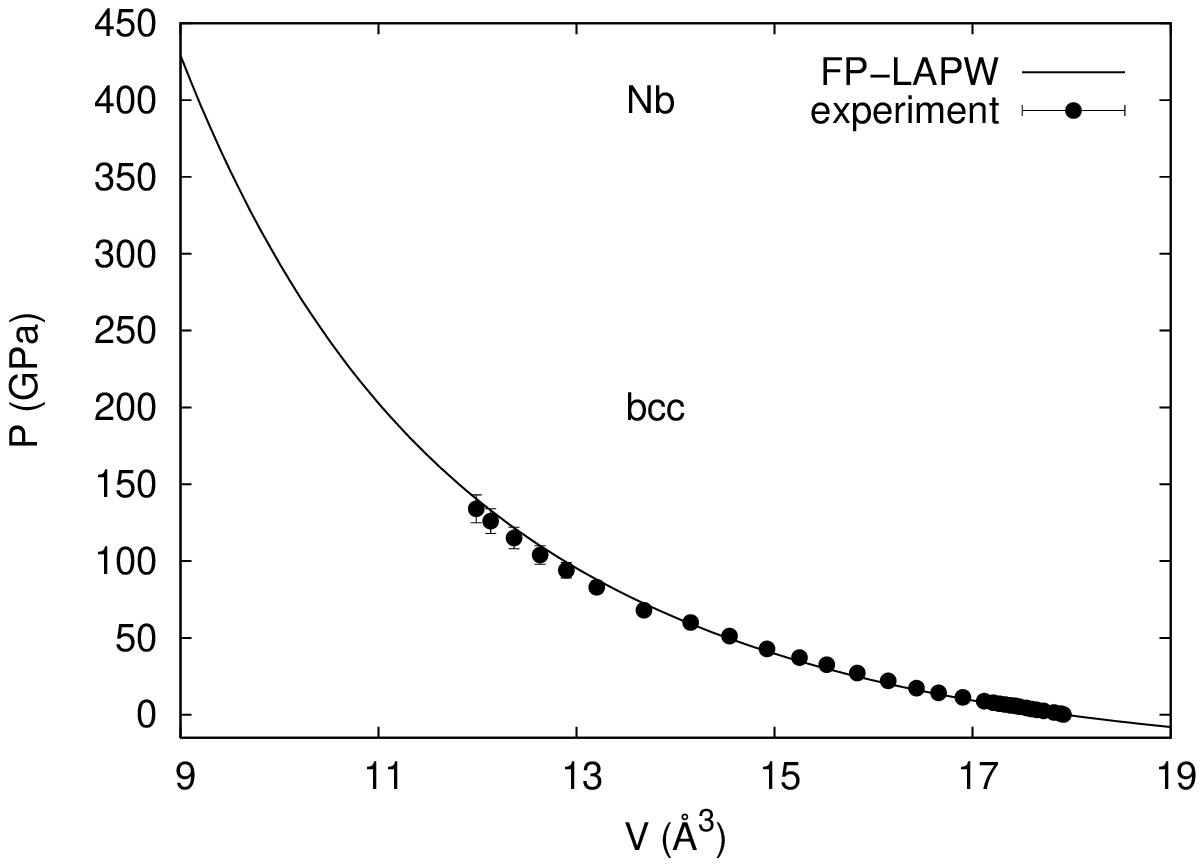} }%
\vspace{0.0cm}
\caption{\emph{Top}: Energy differences $\Delta E$ of the 
$\omega$, hcp and fcc phases with respect to bcc in Nb as function of volume.
\emph{Bottom}: Calculated EOS of Nb compared with 
experimental data of Ref.~[\onlinecite{kenichi06}]~.
}
\label{fig:Nb}
\end{figure}

Experimentally, the observed structure of Nb is bcc over the entire
pressure range from ambient to $\sim 145$~GPa. Our total-energy
results (Fig.~\ref{fig:Nb}) show that hcp, fcc and $\omega$ are all less stable
than bcc by at least $0.2$~eV/atom over the pressure range up to
over $400$~GPa. A possible phase transition at pressures above 400~GPa
may be suggested by the down-turn in the
energy difference between hcp and bcc at $\sim 400$~GPa, but we have not
pursued this possibility. The agreement of our calculated
$P ( V )$ curve with the recent experimental data of 
Kenichi and Singh~\cite{kenichi06} obtained at 
room temperature is excellent. 
Our calculated equilibrium volume $V_{0} = 17.94$~\AA$^3$/atom,
agrees closely with the experimental value at ambient 
conditions~\cite{donohue74} of $17.98$~\AA$^3$/atom.

\subsection{Molybdenum}
\label{sec:Mo}

\begin{figure}
\centering
{\includegraphics[width=0.65\linewidth]{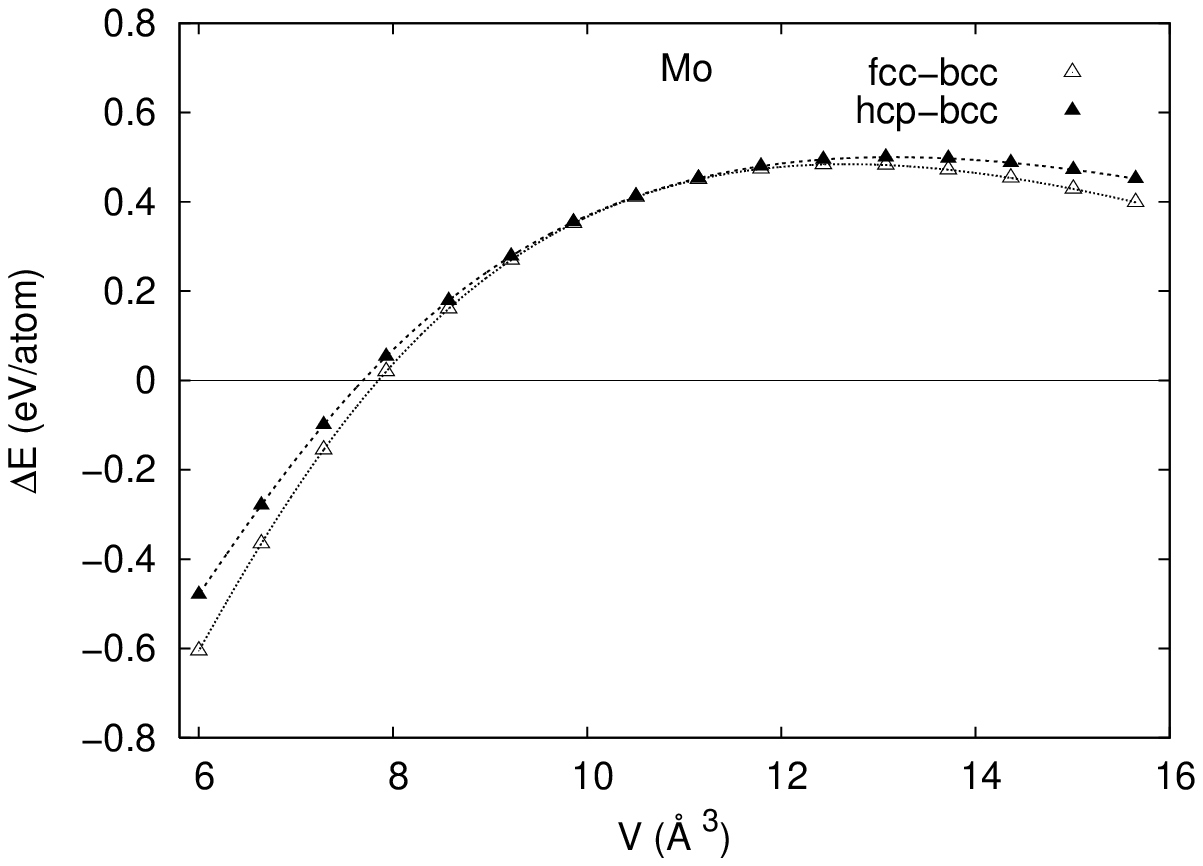} }%
{\includegraphics[width=0.65\linewidth]{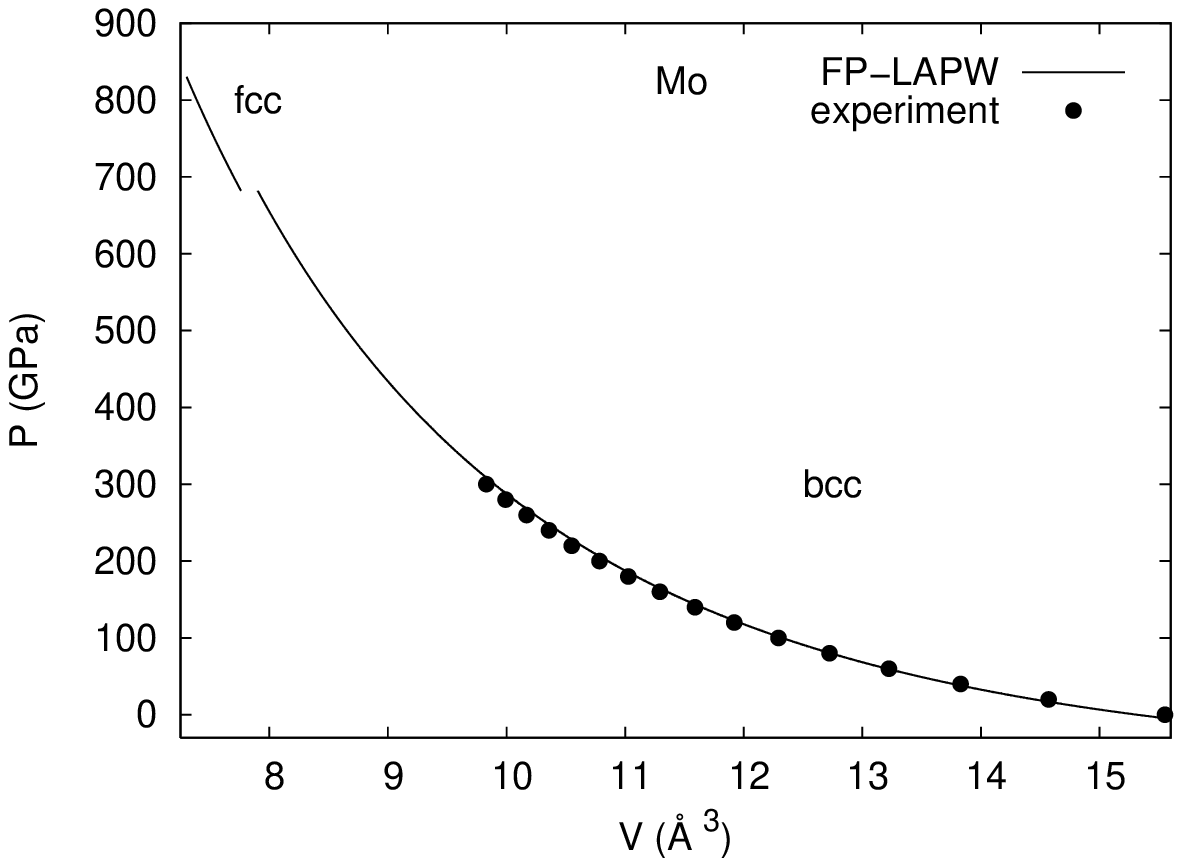} }%
\vspace{0.0cm}
\caption{\emph{Top}: Energy differences $\Delta E$ of the fcc and hcp phases with respect to bcc 
in Mo as function of volume. \emph{Bottom}: Equation of state of Mo 
compared with experimental data.~\cite{hixson92}}
\label{fig:Mo}
\end{figure}

The properties of Mo have been intensively studied both experimentally
and theoretically over the last decade. We mentioned in the Introduction
the continuing controversy over its high-pressure melting curve.
At low temperatures, experiments show that Mo has the bcc structure
for all pressures up to $\sim 400$~GPa. There have been at least
seven previous DFT studies on its phase transitions at higher
pressures, and all agree that there is a transition to either hcp
or fcc, but there is no consensus about 
which high-$P$ form is more stable.
There is also considerable variation of the 
predicted transition pressures.
Our calculated energy differences fcc - bcc and hcp - bcc (Fig.~\ref{fig:Mo})
show clearly that bcc is the stable phase up to a pressure of
nearly $660$~GPa. Above this, both fcc and hcp fall below bcc, but
fcc is below hcp, and the fcc - hcp difference becomes more
negative with increasing pressure. Our calculated $P ( V )$
curve is in excellent agreement with experimental data over the
whole pressure range up to $300$~GPa for which there are data.~\cite{hixson92}

\begin{table}
\begin{center}
\begin{tabular}{l c c c c }
\hline
\hline
   & high-$P$ phase \quad & \quad DFT Method & $E_{\rm xc}$ & $\quad P_{t}$ \\
\hline
 Moriarty~\cite{moriarty92} & hcp & LMTO-ASA & LDA(HL) & 420 \\
 S\"{o}derlind {\em et al.}~\cite{soderlind94} & hcp & 
FP-LMTO & LDA(HL) & 520 \\
 Smirnova {\em et al.}~\cite{smirnova02} & hcp & FP-LMTO & LDA(PZ) & 620 \\
 Jona and Marcus~\cite{jona05} & hcp & FP-LAPW & GGA(?) & 620 \\
\hline
 Christensen {\em et al.}~\cite{christensen95} & fcc & FP-LMTO & 
LDA(BH) & 580 \\
 Boettger~\cite{boettger99} & fcc & LCGTO-FF & LDA(HL) & 660 \\
 Belonoshko {\em et al.}~\cite{belonoshko04} & fcc & PAW & GGA(PW) & 720 \\
 This~work & fcc & FP-LAPW & GGA(WC) & 660 \\
\hline 
\hline
\end{tabular}
\end{center}
\caption{DFT predictions of the high-pressure phase of Mo, with
DFT method and exchange-correlation functional used, and transition
pressure $P_{t}$ (GPa units) from works cited in first column.
Forms of LDA are due to Hedin-Lundqvist~\cite{hedin71} (HL), 
Perdew-Zunger~\cite{perdew82} (PZ) and von Barth-Hedin~\cite{barth72} (BH), and 
of GGA to Perdew-Wang~\cite{perdew91} (PW)
and Wu-Cohen~\cite{wu06} (WC); the form of GGA used by Jona and Marcus
was unspecified.}
\label{tab:Mo}
\end{table}

We summarize in Table~\ref{tab:Mo} all DFT calculations on the
high-$P$ transition in Mo, noting the predicted high-$P$ structure
and transition pressure $P_t$ to that structure, together with
the DFT implementation and exchange-correlation functional. In
general, we believe that work in which careful attention has not
been give to convergence with respect to all technical
parameters and the treatment of semi-core states must be regarded
as less reliable. However, it remains difficult to draw conclusions
about the causes of the very large differences between predictions,
in some cases based on exactly the same exchange-correlation functional.
Further comments will be made in Sec.~\ref{sec:discussion}.

\subsection{Technetium}
\label{sec:Tc}

\begin{figure}
\centering
{ \includegraphics[width=0.65\linewidth]{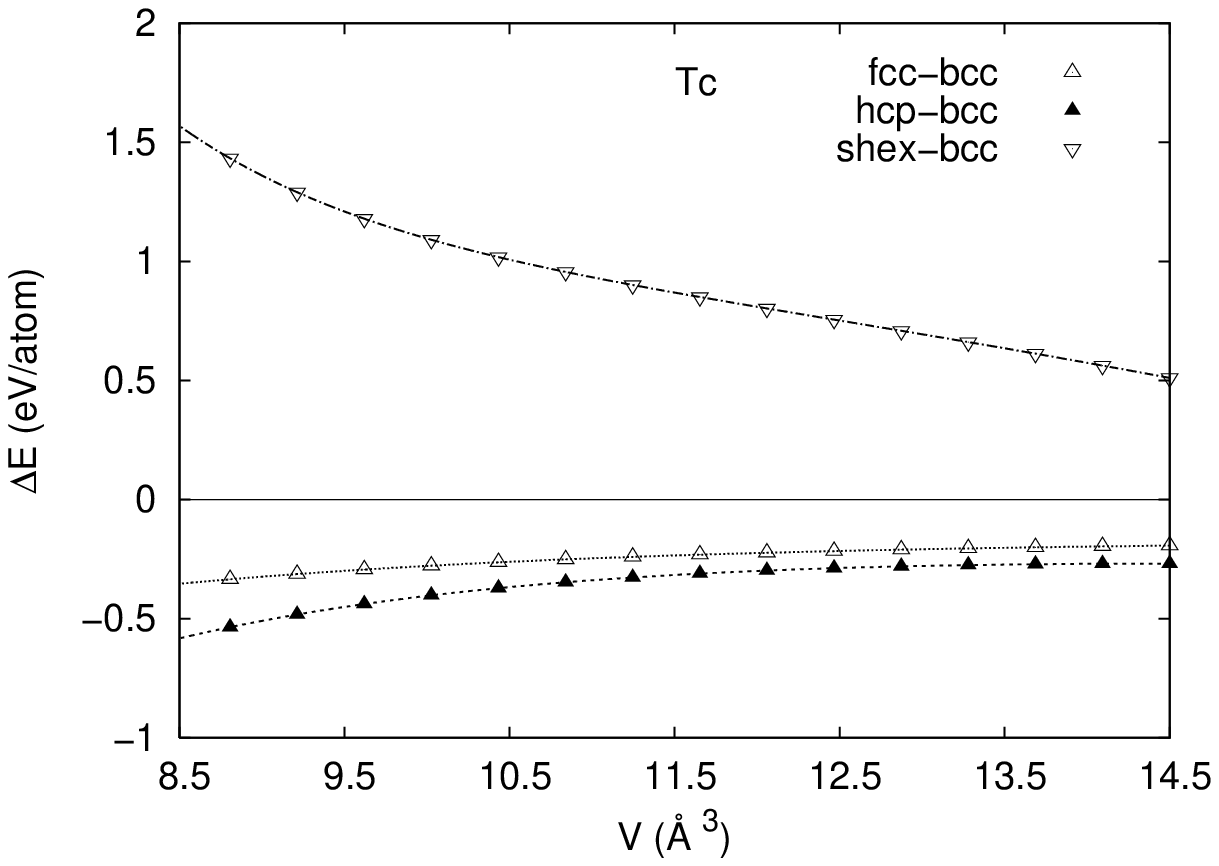} }%
{ \includegraphics[width=0.65\linewidth]{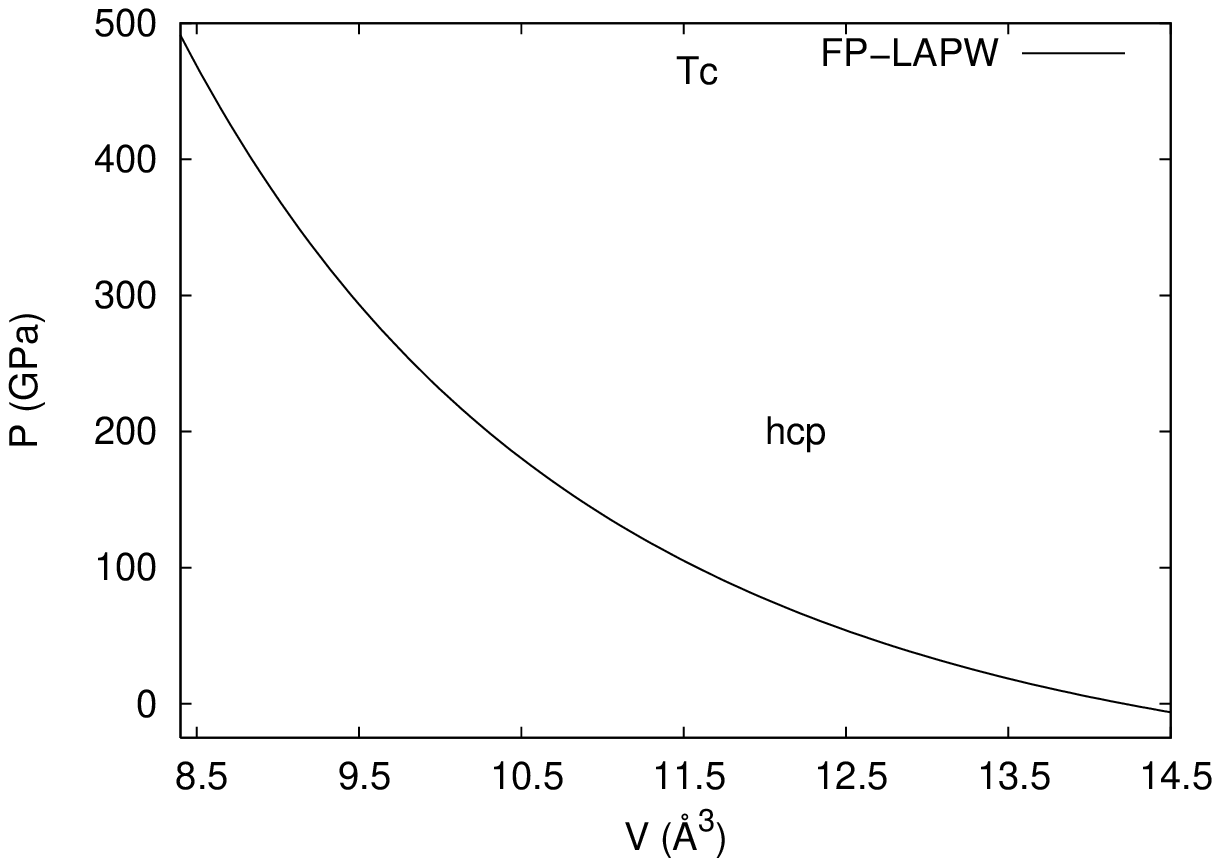} }%
\vspace{0.0cm}
\caption{\emph{Top}: Energy differences $\Delta E$ of the fcc, hcp 
and simple hexagonal (${\rm A_{f}}$) structures with respect to bcc
in Tc as function of volume.
\emph{Bottom}: Equation of state of Tc up to $500$~GPa. 
}
\label{fig:Tc}
\end{figure}

Tc has the hcp structure at ambient pressure, but little seems
to be experimentally known about its behaviour under pressure.
Our results for the energies of the fcc, hcp and simple hexagonal
structures relative to bcc (Fig.~\ref{fig:Tc}) show that hcp is the most
stable up to pressures of at least $500$~GPa, and there is no indication
that any of the other structures will become more stable at higher
pressures than this. The calculated $P ( V )$ curve of hcp Tc
is reported in Fig.~\ref{fig:Tc}, but there appear to be no experimental data
to compare it with. Our value for the equilibrium volume of 
$14.21$~\AA$^{3}$/atom agrees closely with the experimental 
value~\cite{mcmahon06} $14.26$~\AA$^{3}$/atom.

\subsection{Ruthenium}
\label{sec:Ru}

\begin{figure}
\centering
{\includegraphics[width=0.65\linewidth]{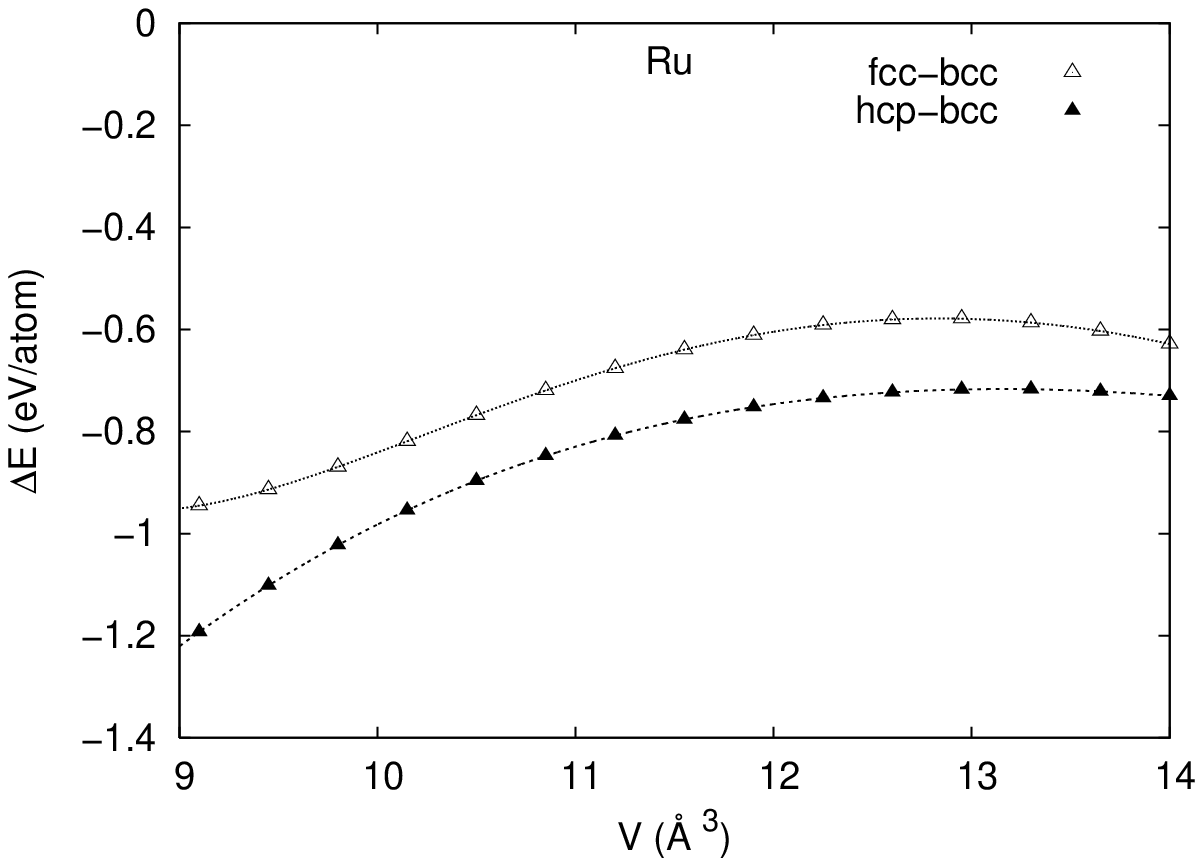} }%
{\includegraphics[width=0.65\linewidth]{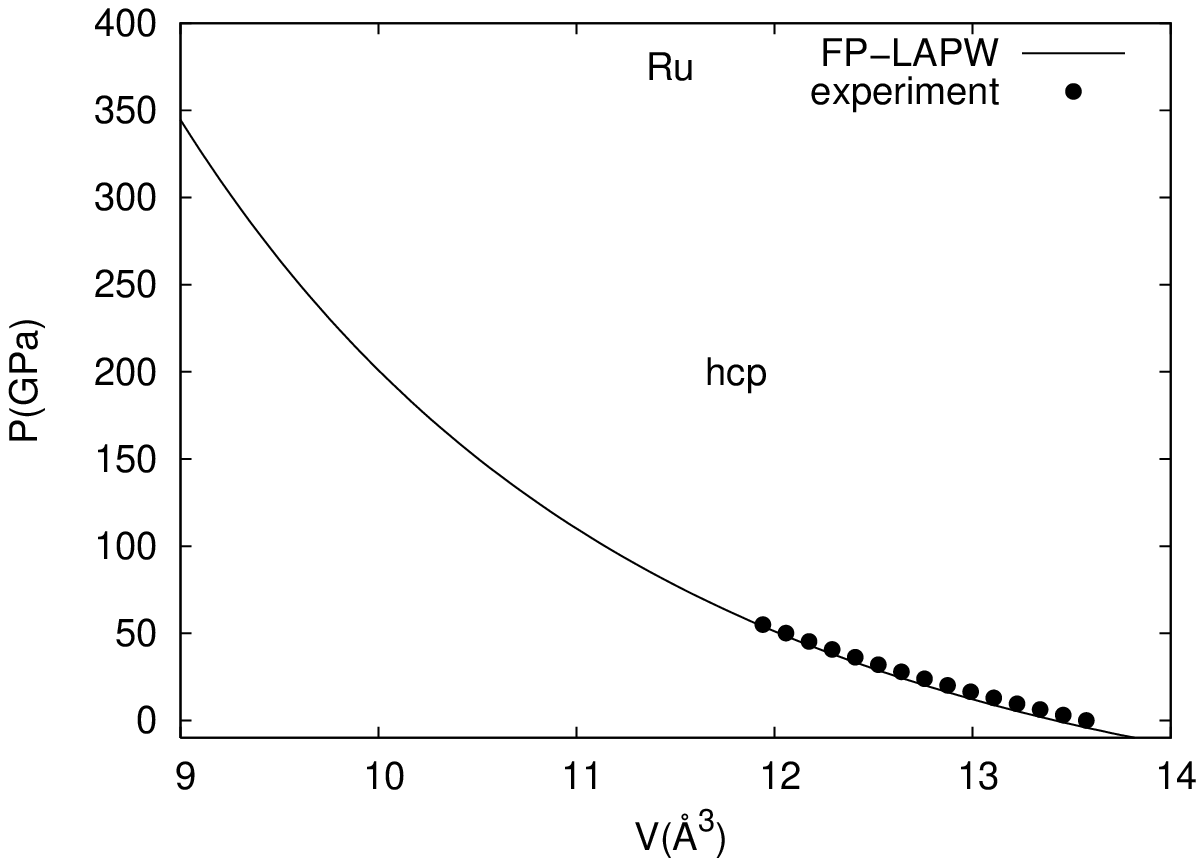} }%
\vspace{0.0cm}
\caption{\emph{Top}: Energy differences $\Delta E$ of the hcp and fcc 
structures with respect to bcc in Ru as function of volume.
\emph{Bottom}: Equation of state of Ru up to $P \sim 400$~GPa 
compared with experimental data.~\cite{cynn02}}
\label{fig:Ru}
\end{figure}

At ambient pressure, the structure of Ru is hcp. Experiments have
been performed up to $56$~GPa, and no phase transition has been found.
Our results for the energies of the fcc and hcp structures relative 
to bcc (Fig.~\ref{fig:Ru}) show that hcp is the most stable up to pressures 
of at least $400$~GPa. 
As can be seen in Fig.~\ref{fig:Ru}, our calculated EOS of hcp Ru 
is in very good agreement with recent experimental data.~\cite{cynn02} 
In particular, we obtain an equilibrium volume of 13.42~\AA$^3$/atom
which is close to the experimental value~\cite{hellwege88,kittel76}
of 13.47~\AA$^3$/atom. (An earlier calculated equilibrium
volume~\cite{finkel71} of 13.57~\AA$^3$/atom is in rather poor
agreement with our value.) We also note that our calculated
$c / a$ value of $1.58$ is exactly the value found experimentally.~\cite{cynn02}

\subsection{Rhodium and Palladium}
\label{sec:RhPd}

\begin{figure}
\centering
       { \includegraphics[width=0.65\linewidth]{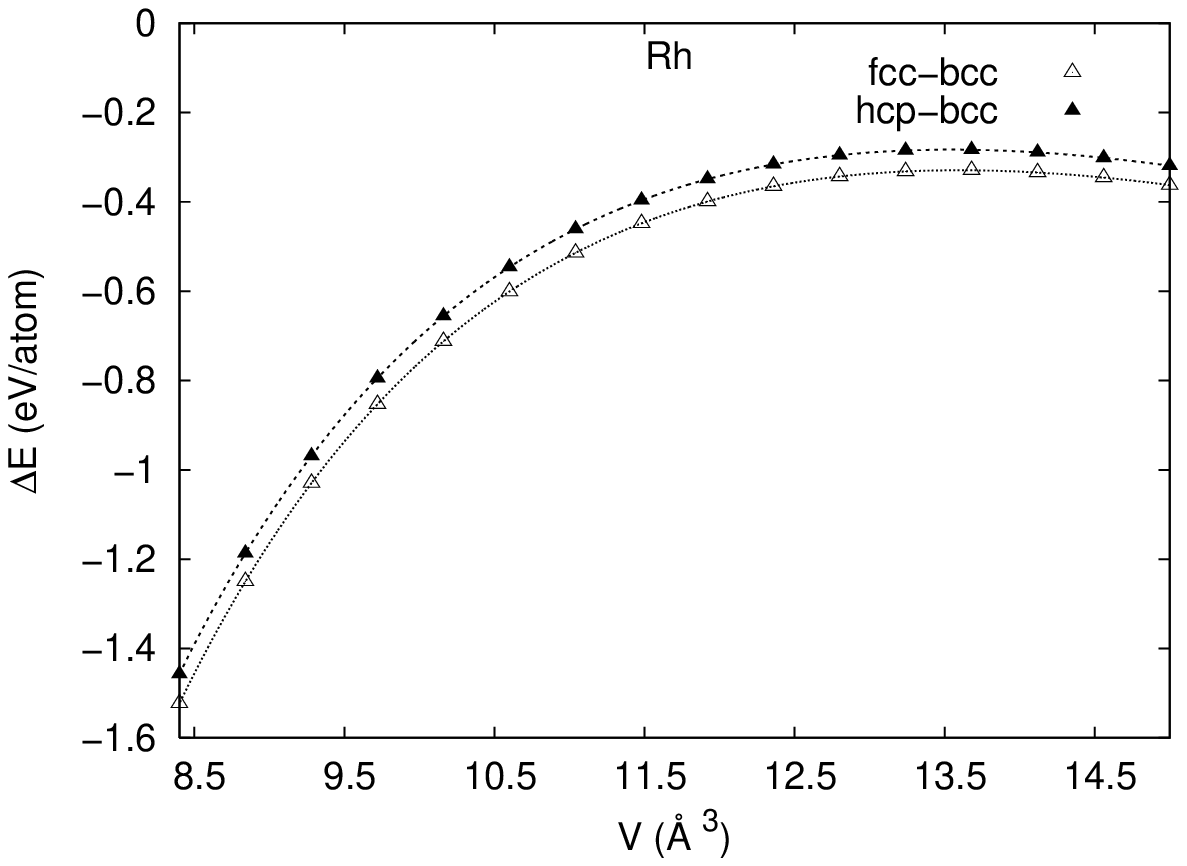} }%
       { \includegraphics[width=0.65\linewidth]{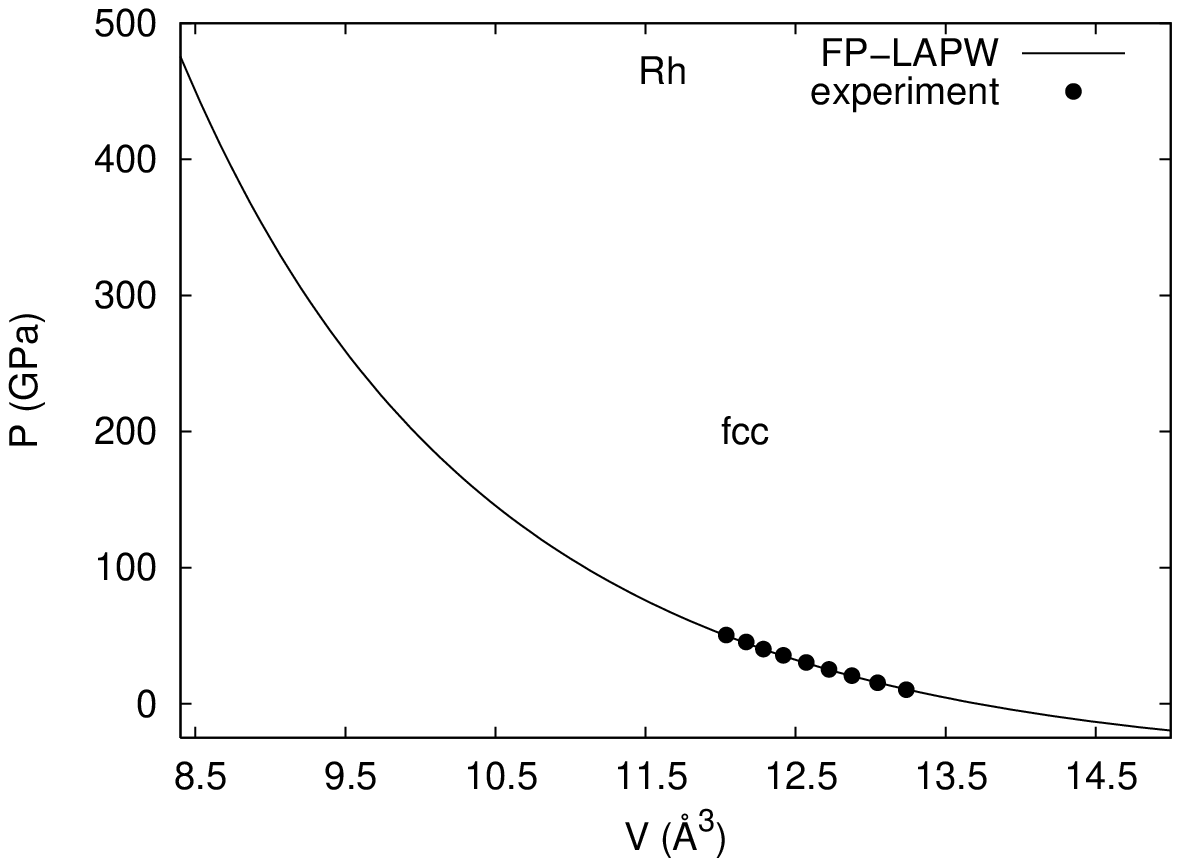} }%
 \vspace{0.0cm}
 \caption{\emph{Top}: Energy differences $\Delta E$ of the fcc and hcp structures with respect to bcc in Rh as function of volume.
	  \emph{Bottom}: Equation of state of Rh up to $P \sim 500$~GPa compared with experimental data.~\cite{rice58}}
\label{fig:Rh}
\end{figure}

\begin{figure}
\centering
       { \includegraphics[width=0.65\linewidth]{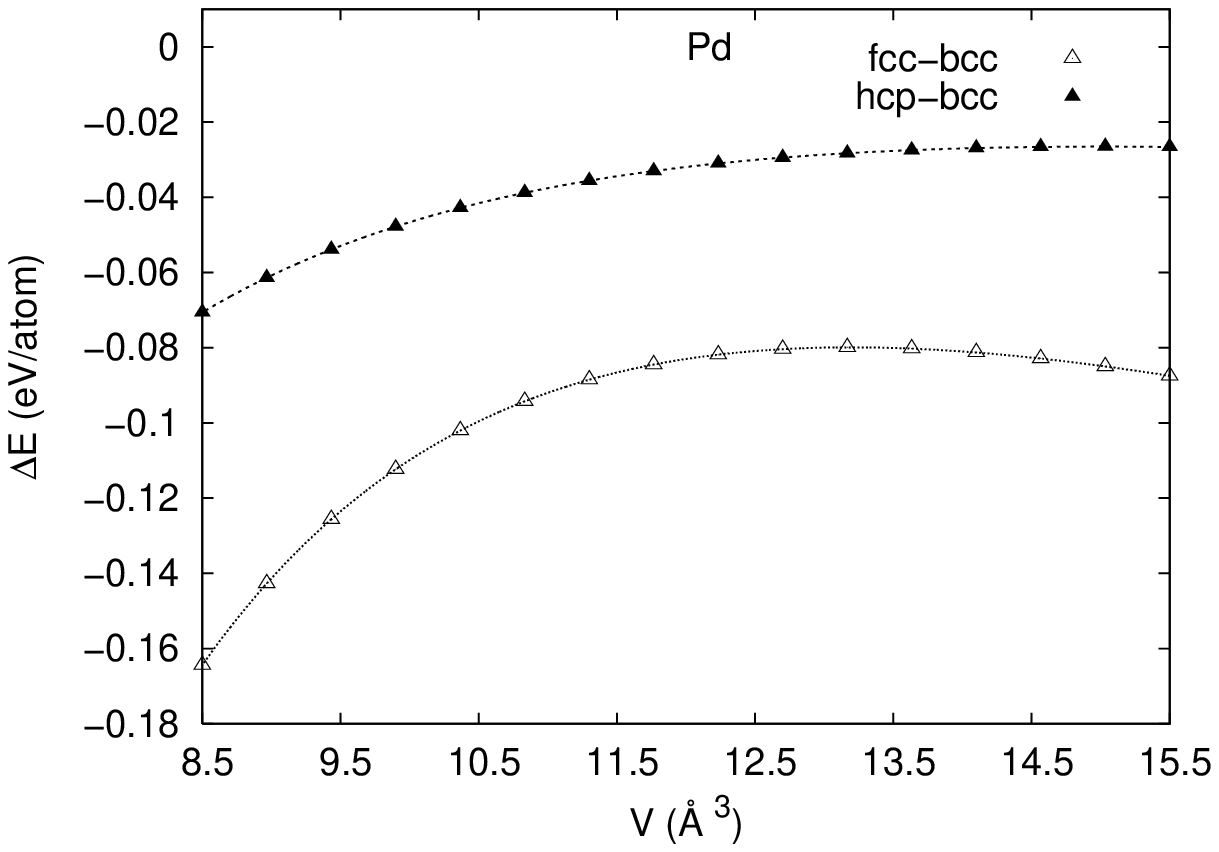} }%
       { \includegraphics[width=0.65\linewidth]{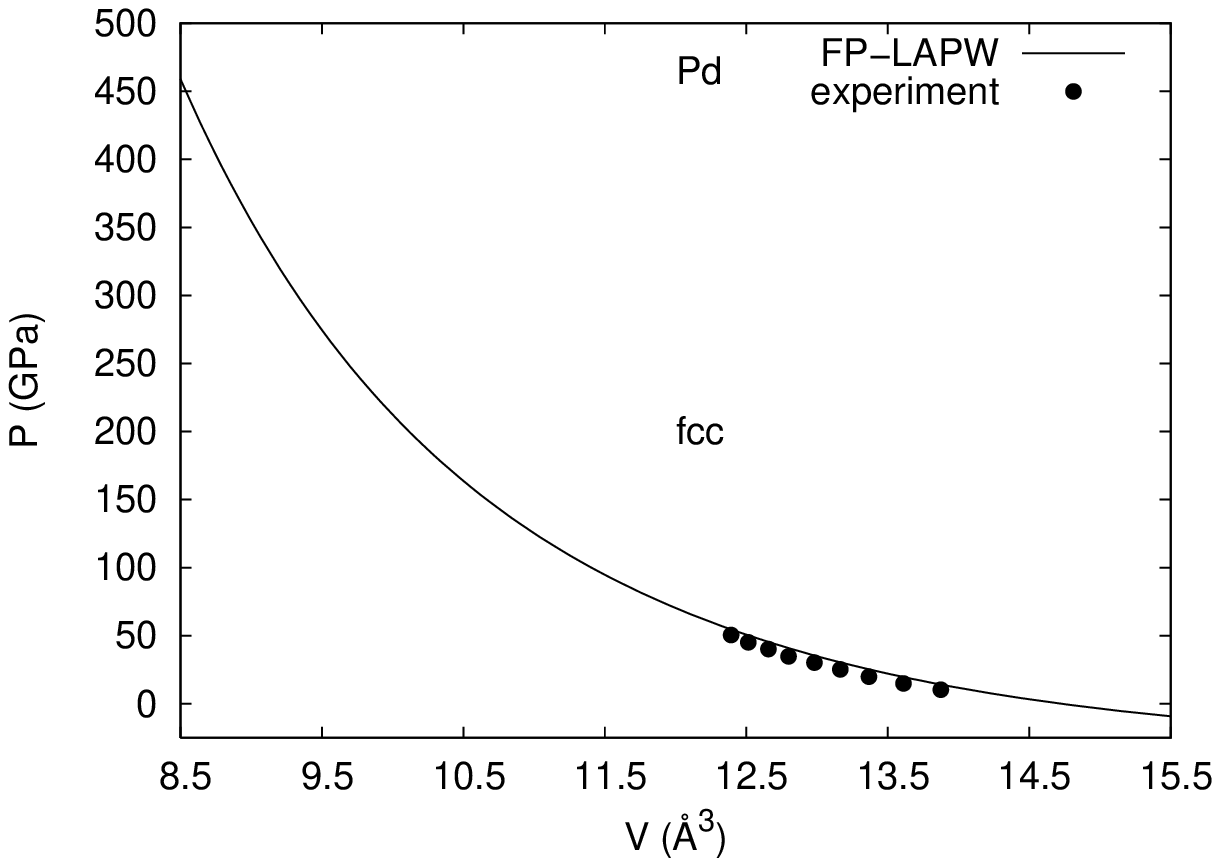} }%
 \vspace{0.0cm}
 \caption{\emph{Top}: Energy differences $\Delta E$ of the fcc and hcp structures with respect to bcc in Pd as function of volume.
          \emph{Bottom}: Equation of state of Pd up to $P \sim 450$~GPa compared with experimental data.~\cite{rice58}
}
\label{fig:Pd}
\end{figure}

The last two elements treated here, Rh and Pd, both have the
fcc structure at ambient pressure, and retain this structure
up to the highest pressures reached so far 
experimentally~\cite{perez64,mao78}
($\sim 50$ and $77.4$~GPa, respectively). 
Our results for the energy differences fcc~- bcc and 
hcp~- bcc (Figs.~\ref{fig:Rh} and ~\ref{fig:Pd}) give
no indication that any phase transition will be found
in the range up to $\sim 500$~GPa. As for most of the other 4$d$ elements,
our calculated EOS results (Figs.~\ref{fig:Rh} and ~\ref{fig:Pd}) are in close
agreement with experiment. Our calculated equilibrium
volumes of 13.71 and 14.73~\AA$^{3}$/atom for 
Rh and Pd respectively agree closely with the
experimental values~\cite{perez64,mao78} of 13.75 and 14.72~\AA$^3$/atom.

\subsection{Zero-temperature phase diagram}
\label{sec:zerophase}

The pressures of the transitions for each element (atomic number $Z$)
reported above can be regarded as points on phase boundaries drawn
in the $( P, Z )$ plane. These boundaries are described
by the dependence of $P$ on $Z$ treated as a continuous, rather
than a discrete variable. In the real world, the elements form
a discrete series, and only integer values of $Z$ are available,
but in DFT theory there is no difficulty in treating Z as continuous.
(In tight-binding theories of transition-metal energetics, it is
common practice to treat the number of electrons per atom as continuous.)
However, to save effort, we have not actually attempted to perform
FP-LAPW calculations for non-integer $Z$, preferring to obtain the
phase boundaries by simple interpolation.

\begin{figure}
\centerline{
        \includegraphics[width=0.8\linewidth]{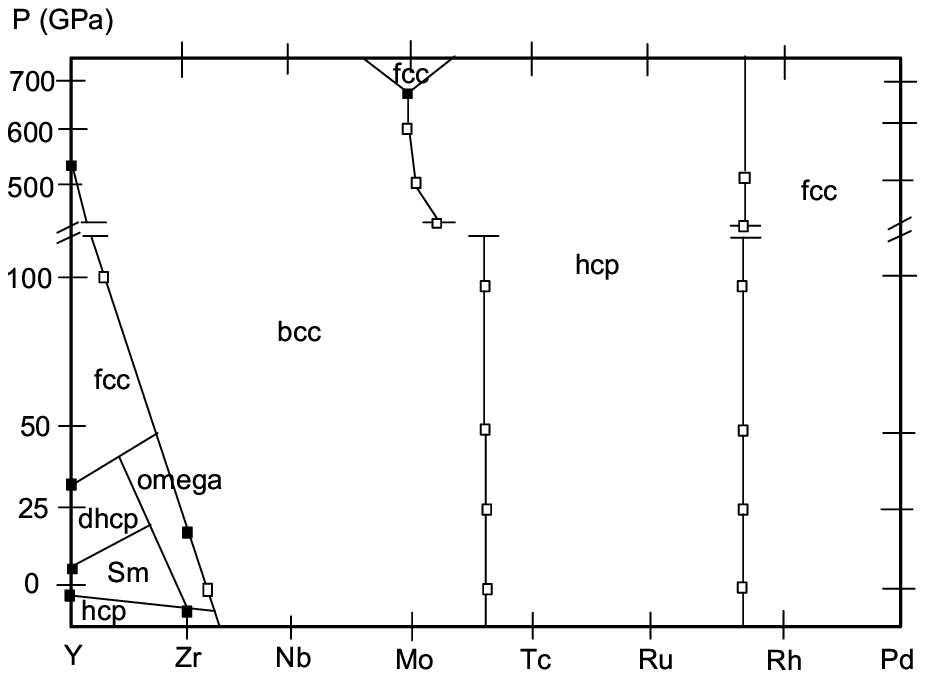}}%
        \caption{Calculated zero-temperature phase diagram of the 4$d$ transition metals series.
		Solid squares correspond to phase transitions obtained directly from the FP-LAPW calculations 
		while open squares show results obtained by interpolation (see text).}
\label{fig:phasediagram}
\end{figure}

To perform the interpolation, we note that at $T = 0$~K the enthalpies 
$H \equiv E + P V$ of coexisting phases must be equal. To take
an example, the hcp~$-$~bcc energy differences for Mo ($Z = 42$) and
Tc ($Z = 43$) at $P = 400$~GPa are $0.62$ and $-0.22$~eV/atom, and the
hcp~$-$~bcc volume differences are $-0.14$ and $-0.12$~\AA$^3$/atom, so that
the hcp~$-$~bcc enthalpy differences are $0.27$ and $-0.52$~eV/atom.
By linear interpolation, we estimate $Z = 42.34$ as the coexistence
value between bcc and hcp at $400$~GPa. We have used this interpolation
scheme to estimate the bcc~$-$~hcp boundary between Mo and Tc,
and the hcp~$-$~fcc boundary between Ru and Rh.
For the complicated region Y~$-$~Zr~$-$Nb, we have performed interpolation only 
for the fcc~$-$~bcc boundary passing through the Y~$-$~Zr region; 
in the other cases, the boundaries have been drawn approximately
by means of straight lines. The resulting generalized phase diagram is
shown in Fig.~\ref{fig:phasediagram}~.

\section{Discussion and conclusions}
\label{sec:discussion}

We noted in the Introduction the general expectation that the
sequence of crystal structures across a transition-metal
series will be the same at high $P$ as at low $P$, but that
the phase boundaries should slope to the left ($d P / d Z < 0$).
The well-known ideas behind this expectation are that
(a)~the sequence of stable structures depends mainly on the
band energy, i.e. the sum of single-electron energies; (b)~the
band energy depends mainly on the $d$-band densities of states
in the different crystal structures, and on the number of
$d$-electrons; (c)~for given atomic number $Z$, increase of
$P$ causes the bottom of the $d$-band to rise with respect to
the $d$-band centroid, so that the effective number of $d$-electrons
increases. Because of this last effect, increase of $P$
should cause a given element to become more like its
neighbour on the right. This is expected to be a stronger
effect for early transition metals, since the increase of 
the effective number of $d$-electrons with compression is 
larger for them.~\cite{pettifor77}

These expectations are confirmed most clearly by the major
boundary separating bcc from more close-packed structures
on the left of the phase diagram. This boundary causes Zr at
rather moderate pressures to adopt the low-$P$ structure
of Nb, and Y to do the same at much higher pressures.
However, the boundary separating bcc and hcp in the middle
of the diagram is almost vertical up to $P \sim 100$~GPa,
and it is only at much higher pressures that it slopes to
the left; furthermore, it appears that Mo never adopts the
low-$P$ structure of its neighbour Tc. Similarly, the boundary
between hcp and fcc on the right of the diagram is practically vertical up
to $P \sim 500$~GPa. 

DFT predictions of the many transitions shown by Y and Zr already
give evidence that DFT provides reasonably good quantitative accuracy
for the pressure dependence of band structures. The bcc-fcc transition
in Mo will probably remain out of reach of experiment for some time.
On the controversial question of whether the high-$P$ transition
in Mo is really bcc-fcc or bcc-hcp, our results support the former.
The reason for this controversy is clearly that the hcp and fcc energies
are very close to each other, so that the calculations have to
be converged to high accuracy to yield reliable conclusions.
We have taken pains to ensure that our prediction of bcc-fcc
is robust with respect to all convergence parameters. Consequently,
we believe that the small and unexpected field of fcc at high $P$
and roughly half band filling is a real effect.

The bcc-fcc transition in Mo is relevant to the interpretation
of the shock experiments mentioned in the Introduction.
When this transition was first studied theoretically,
the predicted transition pressure was $420$~GPa,~\cite{moriarty92} which is much
lower than our value of $660$~GPa. At that time, it seemed likely that
the transition is closely related to the transition seen in
shock experiments~\cite{hixson89} on Mo at $P \simeq 210$~GPa and an (estimated)
temperature of $3100$~K. Our confirmation of earlier work giving
a much higher transition pressure makes it much less likely
that there is a direct connection with the shock transition.
The only way of maintaining this connection would be to postulate
that the transition pressure decreases strongly with increasing
temperature. However, our DFT calculations of phonon frequencies
in the two structures~\cite{cazorlagillan} 
show that the transition pressure actually {\em increases}
with temperature, so that a direct connection with the
shock transition is completely ruled out.

Turning now to detailed comparisons with experiment, we have
shown that all our calculated equations of state $P ( V )$ 
agree very closely with experiment.
The sequences of stable structures with increasing pressure
are always correctly predicted, where experimental data are
available, but our calculations have a clear tendency
to underestimate transition pressures $P_{t}$ by typically
$\sim 10$~GPa. This can be attributed to a relative shift
$\delta E$ of the energies of coexisting phases. Since at
$T = 0$~K, the enthalpies $H$ of coexisting phases are equal,
a shift of $P_t$ is related to $\delta E$ by:
\begin{equation}
\delta E = \delta P_{t} \left( \left( \partial H_{1} / \partial P \right)_{T} -
\left( \partial H_{2} / \partial P \right)_{T} \right) ~,
\end{equation}
where $H_{1} ( P )$ and $H_{2} ( P )$ are the enthalpies of the two phases.
Since $( \partial H / \partial P )_{T} = V$, the shift $\delta P_{t}$
can be estimated as $\delta P_{t} = \delta E / ( V_{1} - V_{2} )$,
where $V_{1}$ and $V_{2}$ are the coexisting volumes. Then a shift
$\delta P_{t} = 10$~GPa equates to a ratio
$\delta E / ( V_{1} - V_{2} ) = 60$~meV/\AA$^3$~atom. Taking an example,
the volume change $V_{1} - V_{2}$ at the hcp-$\omega$ transition
of Zr was calculated to be $0.1$~\AA$^3$/atom, so it would require
a relative shift of 6~meV/atom to account for 
the error in $P_{t}$~. On the other hand,
for the $\omega$-bcc transition of Zr, for which
$V_{1} - V_{2} = 0.6$~\AA$^3$/atom, the relative shift would
have to be $36$~meV/atom. It is easy to see that room-temperature
thermal effects are unlikely to be the cause. Estimating
the vibrational free energy per atom $F_{\rm vib}$ as
$3 k_{\rm B} T \ln ( \hbar \bar{\omega} / k_{\rm B} T )$,
with $\bar{\omega}$ the geometric-mean phonon frequency,
a difference $\delta \bar{\omega}$ between coexisting phases
causes a relative free energy shift 
$\delta F_{\rm vib} \simeq 3 k_{\rm B} T \delta \bar{\omega} / \bar{\omega}$.
With $3 k_{\rm B} T = 80$~meV at room temperature, unreasonably
large $\delta \bar{\omega} / \bar{\omega}$ values would
be required. On the other hand, we have cited evidence (Table~\ref{tab:Zr})
that differences between density functionals can lead to
$P_{t}$ shifts of order $10$~GPa in the $\omega$-bcc transition of Zr.
However, there is a third possible cause, namely errors
of DFT implementation. We have made efforts to ensure that our
FP-LAPW energies are converged to $\sim 1$~meV/atom with
respect to plane-wave and angular-momentum cut-offs, but there remain
possible linearization errors, whose size is difficult to estimate.
It seems clear that implementation errors must be the reason
for some of the very large differences between DFT predictions
of $P_t$ values. We think this explains the difference of over
$200$~GPa in the predicted $P_{t}$ of the fcc-bcc transition in Y
(Table~\ref{tab:Y}), because the earlier DFT work was done at
a time when the treatment of semi-core states was less well
developed. The range of $\sim 300$~GPa in predicted $P_{t}$ values for
the high-$P$ transition in Mo must also be attributed to implementation
errors in early work.

We conclude by recalling that the present calculations
on the zero-temperature $( P, Z )$ phase diagram are intended
as a prelude to the systematic mapping of the $( P, T, Z )$
diagram, including the solid-liquid coexistence surface as a
function of $P$ and $Z$. In spite of major progress in the first-principles
calculation of melting curves over the past 10 years,~\cite{alfe99,alfe02,alfe04,belonoshko04} the
computation of the entire $( P, T, Z )$ phase diagram is clearly
a major challenge, which will need to be tackled in stages.
The use of DFT to calculate harmonic vibration frequencies~\cite{souvatzis08}
will allow quite rapid progress at temperatures up to about one
third of the melting temperature, and we have already
reported systematic calculations of this kind for Fe, Ta and Mo
over a wide range of pressures.~\cite{vocadlo03,cazorla08} For melting curves, several
first-principles methods are available.~\cite{alfe02,alfe04} However, we believe
it will also be valuable to make rapid and more approximate
surveys using tight-binding methods, and we plan to report on
this approach in the near future.

\acknowledgments
The work was supported by EPSRC-GB Grant No. EP/C534360, which
is 50\% funded by DSTL(MOD). The work was conducted as
part of a EURYI scheme award to DA as provided by
EPSRC-GB (see www.esf.org/euryi).


\begin{thebibliography}{99}
\bibitem{dhcp} The dhcp and $\alpha$Sm structures are both 
variations of close packing. Whereas hcp and fcc
have the sequences ABAB... and ABCABC..., dhcp 
goes ABACABAC... and $\alpha$Sm goes ABCBCACAB...
\end{thebibliography}
\end{document}